\documentclass[preprint2]{aastex}

\begin{document}

\title{Optical Detection of Star Formation in a Cold Dust Cloud in the Counterjet Direction of Centaurus A}

\author{William C. Keel\altaffilmark{1}}
\affil{Department of Physics and Astronomy, University of Alabama,
Box 870324, Tuscaloosa, AL 35487}
\email{wkeel@ua.edu}
\and
\author{Julie K. Banfield and Anne Medling\altaffilmark{2,3}}
\affil{Research School of Astronomy and Astrophysics, Australian National University,
Canberra, ACT 2611, Australia}
\and
\author{Susan G. Neff}
\affil{NASA Goddard Space Flight Center}

\altaffiltext{1}{SARA Observatory}
\altaffiltext{2}{Current address: Cahill Center for Astronomy \& Astrophysics, California Institute of Technology, MS 249-17, Pasadena, CA 91125, USA}
\altaffiltext{3}{Hubble Fellow}

\begin{abstract}
We have identified a set of optical emission-line features 700\arcsec  (12 kpc) to the southwest of the nucleus of Centaurus A, roughly opposite to
the radio jet and well-known optical emission filaments associated with the northern radio structure (along the
axis of the southwestern radio lobes, although there is no coherent counterjet at this radius).
We use integral-field optical spectroscopy to trace the ratios of strong emission lines, showing
changes in excitation across the region, and significant local reddening. The emission regions are spatially associated with far-infrared emission peaks in one of two
cold dust clouds identified using {\it Herschel} and {\it Spitzer} data, and there may be a mismatch between the
low temperature of the dust and the expected heating effect of young stars.  The strong emission lines have ratios consistent with photoionization 
in normal H II regions, requiring only modest numbers of OB stars; these stars and their cooler accompanying populations
must be obscured along our line of sight. These data fit with a picture of fairly ordinary formation of clusters in
a large giant molecular cloud, or network of such clouds. The location, projected near the radio-source axis and 
within the radius where a starburst wind has been inferred on the other side of the galaxy, raises the question
of whether this star-forming episode was enhanced or indeed triggered by an outflow from the central parts of
Centaurus A. However, optical
emission-line ratios and line widths limit the role of shocks on the gas, so any
interaction with an outflow, associated either with the radio source or star formation
in the gas-rich disk of Centaurus A, can at most have compressed the gas weakly.
We speculate that the presence of similar star-forming regions on both sides of the galaxy, contrasted with the
difference in the character of the emission-line clouds,
reflects the presence of a collimated radio jet to the northeast and perhaps anisotropic escape of ionizing 
radiation from the AGN as well. In this view, the star formation on the southwestern side of Cen A could be enhanced by a broad outflow
(whether originated by a starburst or AGN), distinct from the radio jet and lobes.
\end{abstract}


\section{Introduction}

Centaurus A (NGC 5128), as the nearest galaxy hosting a large double radio source, has long
played a special role in our understanding of similar objects. It displays
a host of characteristic features observable in unique detail - a large-scale
double source spanning 600 kpc, radio and X-ray jets on scales
up to 3 kpc, merger signatures in stars, gas, and dust, and
optical emission-line features often attributed to interaction between
the propagating jet and ambient interstellar medium, generating both
shock ionization and star formation. 

Optical emission regions and possible young stars near the northeastern radio jet were identified by 
\cite{Blanco75}; spectroscopy by \cite{Osmer78} confirmed that both normal H II region and bright supergiants
were present. \cite{GrahamPrice} used higher-resolution spectra to show that large ``turbulent" velocities in the range
400 km s$^{-1}$ appear across few-arcsecond scales, well beyond what could be produced in star-forming regions.
Data presented by \cite{PDC75} show that multiple ionizing mechanisms may be at work, with line ratios in some
parts of the first-discovered emission regions have [O I] too strong for ordinary H II regions ionized by stars.
The ionization of the gas in these filaments is complex; \cite{Morganti91} suggest a dominant role for photoionization 
by the beamed continuum of a small scale jet. In contrast, \cite{Sutherland} present models showing that 
the small-scale velocity structure matches shock excitation of the spectral lines; {\it Chandra} X-ray data analyzed by \cite{EvansKoratkar} support this through showing very hot gas along one side of the most prominent filament. \cite{Rejkuba} used color-magnitude diagrams to identify
local young supergiant stars; comparison of their locations to the ionized gas structure provides additional evidence for
multiple ionizing mechanisms, while a deeper {\it Hubble Space Telescope} analysis by \cite{Crockett} makes this case more
stringently in the brightest emission region.

\cite{Santoro2015} show that the dynamics of ionized gas in the prominent filaments are consistent with the ambient H I clouds, including
components in regular rotation about the galaxy as well as being entrained by the radio jet. Additional MUSE observations
by \cite{Santoro2016} show that even on small scales in these filaments, there is a changing mix of ionization
mechanisms, with
embedded star formation and photoionization by the distant AGN both indicated by emission-line ratios. 
\cite{McKinley2017} suggest that one of the outermost emission-line filaments may result from interaction with a wind rather that a collimated
jet, and speculate that the southwestern jet may not intersect suitable cold gas to produce similar effects in the ``counterjet"
direction.

As part of a study of these
emission features, we have found an emission-line feature along the
``counterjet" direction, which seems to have been previously 
unremarked\footnote{As this study was in progress, we found that the emission feature 
appears in the long-exposure, very deep composite image presented by
Rolf Olsen at
http://www.rolfolsenastrophotography.com/Astrophotography/Centaurus-A-Extreme-Deep-Field/
and is described in his text as possibly related to the jet:
``A corresponding faint trace of nebulosity, likely related to the otherwise invisible Southern jet, 
is also noticeable as a small red smudge on the opposite side of the galaxy core".}.
We present here morphological and spectroscopic observations of this object,
which appears to consist of (possibly multiple) H II regions powered by young stars. The
H$\alpha$ emission is spatially
coincident with the peaks of the cold dust feature found
by \cite{Auld}  from
{\it Herschel} observations, furnishing an interesting puzzle as to how 
a star-forming region coexists with the dust without generating
a higher temperature than observed.

The radio structure does not have a distinct counterpart to the northern
jet on the southern side at this distance. The regions we observe
fall in a minimum in radio flux between inner and outer lobes 
\citep{Junkes}, making direct interaction with a jet
unlikely.

In computing sizes and luminosities, we adopt a distance 3.7 Mpc (scale 17.9 
pc arcsecond$^{-1}$), following the Cepheid results of \cite{Tully2013} 
and the red-giant studies from, e.g., \cite{Crnojevic} and
\cite{Tully2015}.

\section{Observations}

\subsection{Identification and optical imaging}

\begin{figure*} 
\includegraphics[width=130.mm,angle=270]{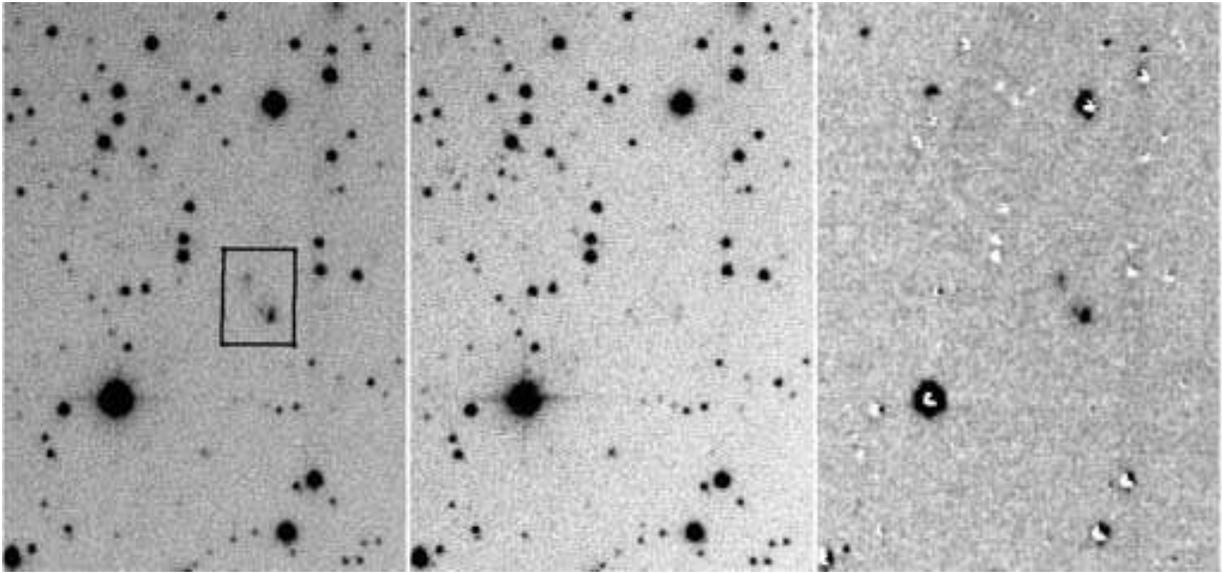} 
\caption{SARA H$\alpha$+[N II] (left) and R (center)  images of the southwestern field in Centaurus A. The third 
panel shows a continuum-subtracted version,
smoothed with a 3-pixel (1.2\arcsec) Gaussian;
approximate PSF matching was done by convolving the R images with a Gaussian 
of 1.0\arcsec\ FWHM, but residuals appear in the cores of bright stars.
The boxed features are essentially pure line
emission. Two starlike objects to their northeast have
significant residual flux in H$\alpha$, but may be foreground stars. Each panel spans
$178 \times 266$\arcsec, with north at the top. The nucleus of
Centaurus A lies about 700\arcsec\ to the upper left. 
Faint diffuse streaks to the north in the
H$\alpha$ and difference images are scattered light from a 9th-magnitude
star.}
 \label{fig-halphamontage}
\end{figure*}

The emission region was identified in May 2014 using an H$\alpha$ filter
of FWHM 75 \AA\ on the remotely-operated SARA 0.6m telescope
at Cerro Tololo, Chile \citep{SARA}. A CCD system from ARC of San Diego
operated at -110 C; the pixel scale was 0.38\arcsec.
We coadded images totaling 7 hours' integration 
in this filter, and 70 minutes of continuum imaging in the R band, 
obtained between May and August 2014. 
Flux calibration used \cite{Landolt} standard stars, carried to the
narrow filter using the ratio of filter widths. The 
narrowband image
(Fig. \ref{fig-halphamontage}) shows a set of diffuse H$\alpha$ emission 
regions, and two starlike
objects with strong residual H$\alpha$ flux after continuum subtraction
using the $R$ image. Color terms in this subtraction will be modest, 
because H$\alpha$ is near the center of the $R$ band. The image gives a 
total H$\alpha$+[N II] flux within the brightest $15 \times 15$\arcsec\ region
of $6.1 \times 10^{-15}$ erg cm$^{-2}$ s$^{-1}$ \AA$^{-1}$, about 50\% lower than 
implied by the flux calibration of the integral-field spectroscopic data.

For comparison, we also make use of a similar H$\alpha$/$R$ image pair centered on the core
of Centaurus A, summing total exposures of 4 hours in H$\alpha$ and 50 minutes in $R$, and
H$\alpha$ and $I$ images of a location in the northeast emission filaments obtained in February 1987
using the ESO/MPI 2.2m telescope at La Silla, as described by \cite{Keel89}

Coordinates were derived using the astrometry.net Web service
\citep{Lang}, automatically matching
field stars to coordinate catalogs. The emission-line features span
an extreme length of 26\arcsec\ (460 pc) in projection; J2000
coordinates are given in Table \ref{tbl-coords}. This region lies 12 kpc in projection
from the core of Centaurus A.

\begin{deluxetable}{lcc}
\tablecaption{Component positions \label{tbl-coords}}
\tablewidth{0pt}
\tablehead{
\colhead{Component} & \colhead{$\alpha$ (J2000)} & \colhead{$\delta$ (J2000)} }
\startdata
\sidehead{Discrete components:}
Brightest & 13 24 35.224 & -43 09 09.5 \\    
North  &   13 24 35.186 & -43 09 06.5      \\    
East     & 13 24 35.636 & -43 09 06.9          \\    
Northeast & 13 24 36.192 & -43 08 52.2        \\
\sidehead{Starlike emission objects: }
 & 13 24 41.371 & -43 07 29.83            \\
 & 13 24 44.032  & -43 07 03.7               \\  
\enddata
\end{deluxetable}

\subsection{Optical spectroscopy}

Data cubes in both blue and red grating settings were obtained using the
WiFeS integral-field spectrograph \citep{Dopita2007}
at the 2.3m ANU Advanced Technology Telescope 
at Siding Spring. The field of view spans $25 \times 38$\arcsec,
with 1\arcsec\  sampling. Simultaneous exposures in the blue (3500-5700 \AA)
and red (5400-7000 \AA)
ranges were obtained on 4 March 2016, for 900 seconds.
The field covered the three southern discrete components,
as well as more diffuse emission to their northeast, and the
southern edge of the NE component.

Strong, narrow emission lines appear; the mean heliocentric radial
velocity is $cz=773 \pm 6$ km s$^{-1}$ (internal error) for the brightest region, and 
$759 \pm 21$ km s$^{-1}$ for the fainter one just to its north. These compare
to the consensus systemic velocity 547 km s$^{-1}$ from 
NED. All lines are narrow, close to the instrumental resolution;
measured FWHM values range from 1.3-1.8 \AA.
These observations used the B3000/R7000 grating combination, with nominal
resolutions 1.7 \AA\ in the blue and 0.9 \AA\ at H$\alpha$.

The associated continuum is quite faint; even summed over all spatial pixels in the
brightest knot, the continuum S/N is only 0.9 per 0.77-\AA\ pixel at 5000 \AA\  and 1.2 per
0.44-\AA\ pixel near H$\alpha$. This limits what we can learn about associated starlight.
In measuring the Balmer emission lines, we therefore consider the full range of plausible 
corrections for absorption in young stellar populations. While H$\alpha$ absorption in old
populations is weak, with equivalent widths near 2 \AA\, it can be as strong as 12 \AA\ in
type A stars \citep{JHC}. Corresponding values for H$\beta$ are 4--16 \AA\ . Since the line
emission has large equivalent widths, we include the ranges in these corrections in our
uncertainties on Balmer decrement, other ratios involving these Balmer lines, and reddening.
Error contributions from noise were evaluated from empty continuum regions near various emission lines.
We follow \cite{Santoro2016} in converting from Balmer decrements to reddening values and
H$\alpha$ attenuation, which include both foreground Milky Way and internal contributions.
Although the Milky Way absorption is significant in the direction, with $A(\rm H \alpha ) = 0.25$
magnitude from the results of \cite{SchlaflyFinkbeiner}, internal attenuation is dominant in 
each of the three regions where we can measure the Balmer decrement.

\section{Discussion}

\subsection{Relation to cold dust cloud}

\begin{figure} 
\includegraphics[width=60.mm,angle=270]{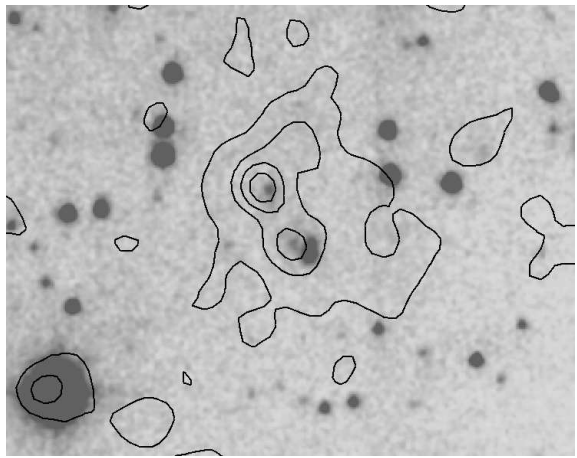} 
\caption{Narrowband H$\alpha$ image as in Fig. \ref{fig-halphamontage}, overlaid with contours
from the 24$\mu$m {\it Spitzer} MIPS observation. Contours are spaced at intervals 0.10 MJy sr$^{-1}$, spaced sparsely
to show the H$\alpha$ peaks. The area shown is $120 \times 150$\arcsec\ with north at the top.}
 \label{fig-IRoverlay}
\end{figure}

Fig. \ref{fig-IRoverlay} overlays the H$\alpha$ on-band image with contours of a {\it Spitzer} $24\mu$m MIPS
observation described by \cite{Brookes}; \cite{Auld} showed that the cloud was clearly detected at this wavelength, where angular
resolution is better than the longer-wavelength data from either {\it Spitzer} or {\it Herschel}. The two
emission-line peaks are closely associated with the 24$\mu$m peak locations. There are no similar H$\alpha$
or 24$\mu$m features within $\approx 300$\arcsec,
leaving little doubt
that these are associated physically rather than only along the line of sight.

\subsection{Ionizing sources and star formation}

\begin{figure} 
\includegraphics[width=80.mm,angle=0]{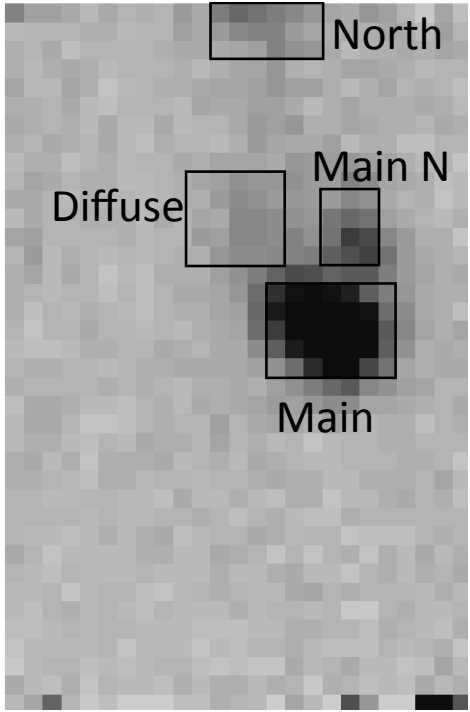} 
\caption{Monochromatic H$\alpha$ image summed across the line's emission profile from the WiFeS data cube, spanning $24 \times 36$\arcsec\ 
with north at the top. The regions summed in Table \ref{tbl-lineratios} are indicated.}
 \label{fig-regions}
\end{figure}

To evaluate likely ionization mechanisms, we measured emission-line ratios
in four spatial regions (Fig. \ref{fig-regions}), fitting Gaussian profiles and linear
baselines, with results given in Table
\ref{tbl-lineratios}. These regions, selected by position and surface brightness, differ significantly in
Balmer decrement H$\alpha$/H$\beta$.

All the [S II] line ratio values cluster near the low-density limit 
of I($\lambda 6717$)/I($\lambda 6731$)=1.43 (mean of all values $1.49 \pm 0.06$).

\begin{deluxetable}{lcccc}
\tablecaption{Emission-line properties \label{tbl-lineratios}}
\tablewidth{0pt}
\tablehead{
\colhead{Quantity} & \colhead{Main} & \colhead{Main N} & \colhead{Diffuse} & \colhead {North} }
\startdata
~F(H$\alpha$) (erg cm$^{-2}$ s$^{-1}$) & $9.3 \times 10^{-15}$ &  $1.3 \times 10^{-15}$ & $1.1 \times 10^{-15}$ & $1.0 \times 10^{-15}$ \\
~H$\alpha$ EW (\AA\ ) & $156 \pm 6$ & $319 \pm 10$ & $58 \pm 9$ & $780 \pm 300$ \\
~[N II] $\lambda 6583$/H$\alpha$ & $0.28 \pm 0.04$ & $0.36 \pm 0.06$ & $0.40 \pm 0.13$ & $0.44 \pm 0.18$ \\
~[S II] $\lambda 6717/ \lambda 6731$ &  $1.53 \pm 0.13$ & $1.52 \pm 0.28$ & $1.52 \pm 0.18$ & $2.0 \pm 0.3$ \\
~[S II] $\lambda 6717+6731$/H$\alpha$ & $0.18 \pm 0.02$ & $0.27 \pm 0.07$ & $0.49 \pm 0.10$ & $0.25 \pm 0.07$ \\ 
~[O I] $\lambda 6300$/H$\alpha$ & 	 $0.025 \pm 0.006$  & $0.018 \pm 0.002$ & ...  & ... \\
~[O III] $\lambda 5007$/H$\beta$ & 	$0.69 \pm 0.10$ & $0.60 \pm 0.30$ &  $0.47 \pm 0.14 $ & 	...  \\
~[O II] $\lambda 3726+3729$/[O III] $\lambda 5007$ & $1.6 \pm 0.5$ &  $< 1.0$ & $4.3 \pm 1.3$  & 	...  \\
~H$\alpha$/H$\beta$ & $7.0 \pm 0.8$ & $10.8 \pm 2.7$ & $5.1 \pm 0.9$ & ... \\
~E$_{B-V}$ & $0.84 \pm 0.11$ & $1.25 \pm 0.23$ & $0.54 \pm 0.1$ & ... \\
~A(H$\alpha$) & $2.10 \pm 0.18$ & $3.14 \pm 0.51$ & $1.35 \pm 0.40$ & ... \\
~L(H$\alpha$) (erg s$^{-1}$) & $1.1\times 10^{38}$ & $3.8 \times 10^{37}$ & $5.7 \times 10^{36}$ & $ > 2.0 \times 10^{36}$ \\
\enddata
\end{deluxetable}

\begin{figure*} 
\includegraphics[trim= 1.5cm 1.0cm 9cm 23cm, clip=true,width=180.mm,angle=0]{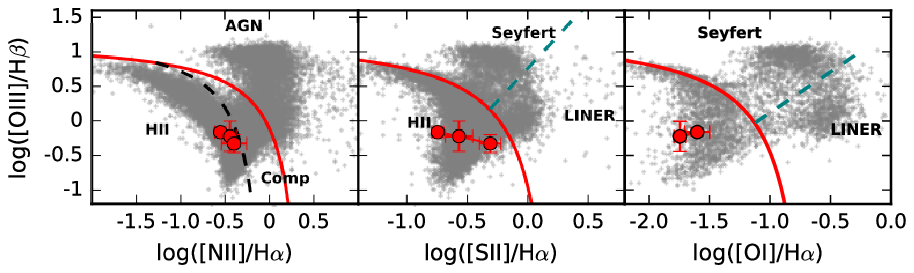} 
\caption{ Emission-line diagnostic diagrams from \cite{BPT} and \cite{VO1987}.  Panels show log([O III]/H$\beta$) vs. (left to right) log([N II]/H$\alpha$), log([S II]/H$\alpha$), and log([O I]/H$\alpha$), with classification boundaries from \cite{Kewley}.  Grey points show all SDSS DR12 galaxies with strong emission lines (SNR$>5$; \citealt{Thomas2013}).  Red points show line ratios of the measured regions listed in Table 2.}
 \label{fig-BPTplots}
\end{figure*}

The location of the regions in the BPT line-ratio diagrams \citep{BPT}, using the revised dividing curves from
\cite{Kewley}, classifies all of them as photoionized by hot stars (Fig. \ref{fig-BPTplots}). This fits with the narrow line widths measured from the
WiFeS data, $\leq 50$ km s$^{-1}$, indicating that shocks fast enough to add significantly to the ionization levels do not
add significantly to the overall ionization level.

These emission regions coincide spatially with the cold dust cloud seen in
{\it Herschel} data by \cite{Auld}, who consider limits on the star formation set
from the dust temperature and (lack of) associated UV sources. Their highest
allowed SFR is set from the 24$\mu$m flux, 0.00012 M$_\odot$ year$^{-1}$. The
GALEX-based near- and far-UV limits are much lower
($<2 \times 10^{-4}$ and $< 5 \times 10^{-5}$
M$\odot$ year$^{-1}$ respectively), indicating that any
associated population of young stars must be obscured from our point of view.

As a star-forming region, the line emission we detect suggests that this object is modest in scale,
with an H$\alpha$ luminosity close to $10^{38}$ ergs s$^{-1}$. This is a few times greater than
that of the Orion Nebula M42, requiring only a few ionizing stars ($< 10$). To compare
with star-formation rates in other environments, we follow \cite{Kennicutt2007}, using
estimating SFR from a linear combination of H$\alpha$ and 24$\mu$m luminosities, as calibrated for disk H II
regions in M51. For the entire southwest complex in Centaurus A, this conversion gives a modest increase 
in effective H$\alpha$ luminosity, and a total SFR $1.3 \times 10^{-3}$ M$_\odot$ year$^{-1}$(for a Salpeter initial-mass function).
This SFR is an order of magnitude greater than the 24$\mu$m limit from \cite{Auld}, and correspondingly larger than their
UV limits. This difference makes sense if the ionizing stars (presumably in clusters) are largely obscured along our line of sight,
but the 24$\mu$m flux is still interestingly low to be associated with even a few ionizing stars. One can, for example, picture
a geometry in which the dust blocks optical and UV light over a small solid angle about the stars, but still hides them from our
direction.

For individual star-forming regions, the relation between long-term star-formation
rate and tracers of massive stars, such as H$\alpha$, has a strong stochastic
element. For example, using the stellar-atmosphere results from
\cite{Vacca}
and \cite{Sternberg}
as in \cite{GildePaz}
shows that the expected H$\alpha$ luminosities from nebulae completely encompassing
the star range from $2 \times 10^{36}$ erg s$^{-1}$ at spectral type B0 to
$10^{38}$ erg s$^{-1}$ at O3. The entire ionizing flux in this complex could
be provided by the equivalent of 9 O7 stars, so small-number statistics would
change the emission-line output strongly as individual stars are formed and evolve.
Even so, there may be more to learn; the limits on IR emission from \cite{Auld} are quite low in comparison to the
SFR rate inferred from H$\alpha$ emission.

The associated dust cloud matches one of the two regions in the H I ``shells" of Centaurus A where \cite{Charmandaris} detected
CO emission, implying a typical H$_2$/H I ratio near unity and consistent with conditions for star formation in the inner regions
of spirals. With an estimated H$_2$ mass of $2 \times 10^7$ M$_\odot$  and linear scale $\approx 0.5$ kpc, this
would be either an exceptionally large giant molecular cloud or a collection of more usual clouds (noting that 
geometrically it may not be easy to distinguish these cases). Under these conditions, it would be common for
parts of a ``blister" H II region to be highly obscured in the optical unless viewed face-on, which fits with the
low UV limits on radiation from young stars.

\begin{figure} 
\includegraphics[width=60.mm,angle=270]{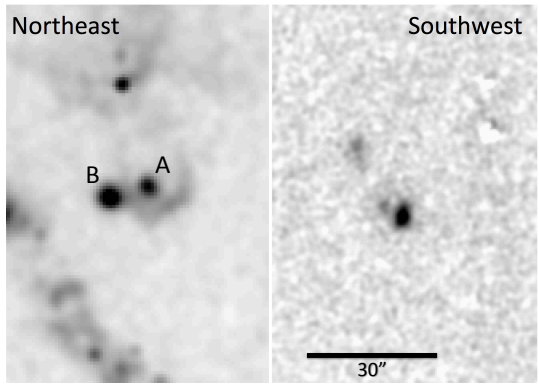} 
\caption{Comparison of the newly observed star-forming regions (right panel) with the H II regions A and B found in the
northeastern emission-line jet by \cite{Santoro2016} (left). The additional filamentary features on the northwest side are ionized by the AGN,
either photoionized or via interaction with the radio jet. Both panels show H$\alpha$ emission and are at the same angular and intensity scales.
The jet region was imaged using the ESO/MPI 2.2m telescope as described by \cite{Keel89}. }
 \label{fig-Northeast-Southwest}
\end{figure}

In H$\alpha$ luminosity and Balmer decrement, these star forming regions are similar to those found embedded in the
northeastern filaments (in what is sometimes known as the ``necklace" structure) by \cite{Santoro2016}. 
In comparison with the  \cite{Santoro2016} regions, these are comparable in scale and observed
luminosity, although perhaps more luminous when dereddened (Fig. \ref{fig-Northeast-Southwest}). We might
speculate that in both cases, interaction with a central outflow \cite{Neff2015} has compressed ambient H I to trigger star formation, but that the
southwestern structure lacks
the additional ingredients of a direct view of the AGN and a collimated radio jet which enhance both effects in 
the gas kinematics and ionization on the ``jet" side.

\section{Summary}

We have described a set of optical emission-line regions found 12 kpc to the southwest of the nucleus of Centaurus A, closely 
coincident with dust and gas structures previously reported. The emission-line ratios as well as
UV and FIR properties are well accommodated as a set of normal H II regions, photoionized by
only a few OB stars.

Even in a system as disturbed as Centaurus A, this is a distant place to find apparently normal H II regions.
Fig. \label{fig-halphanucblob} shows the location superimposed on an H$\alpha$ image of the inner regions, where the
rich distribution of star-forming regions is strongly confined to the warped remnant disk, within a radius of 
212\arcsec \ (3.8 kpc). \cite{Charmandaris} suggest that an outflow from the central regions of the system
has played a role in compressing gas (possibly concentrated dynamically in a way similar to the
stellar shells) so as to trigger such distant star formation. 
On this basis, we might speculate that star formation can be triggered by a broad outflow (driven either by the AGN or
strong star formation in the inner disk), but the southwest region lacks the additional factors producing the
rich optical emission along the north jet. This difference could mean that there is no collimated radio jet on this side, that the
southwestern cloud of gas and dust is not in the right place to be ionized by AGN radiation, 
or that the AGN radiation does not escape effectively on this side.

\begin{figure*} 
\includegraphics[width=130.mm,angle=270]{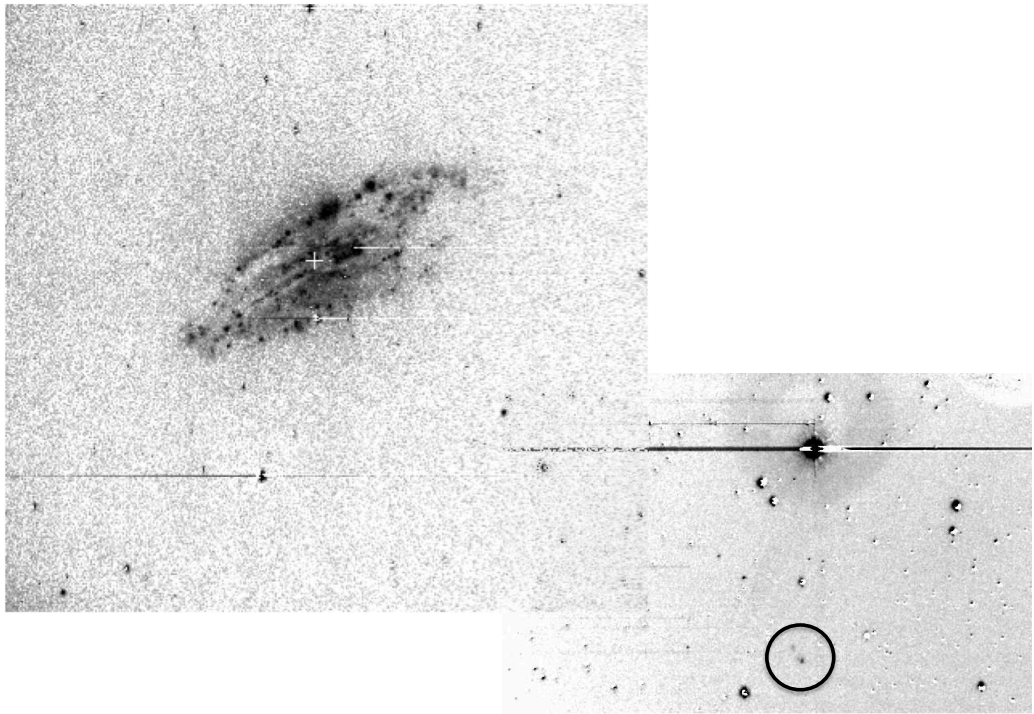} 
\caption{SARA H$\alpha$+[N II] image mosaic showing the newly described H II regions in a wider context
including the core of Centaurus A. The continuum has been subtracted using an $R$-band image, leaving some residuals 
around bright stars. The prominent disk of emission around the nucleus extends to a radius 212\arcsec, while the 
southwest knots are 700\arcsec away.}
 \label{fig-halphanucblob}
\end{figure*} 

\acknowledgments
The IFU spectroscopy was facilitated by a Twitter interaction between two of the authors. 
This research has made use of the NASA/ IPAC Infrared Science Archive, which is operated by the Jet Propulsion Laboratory, 
California Institute of Technology, under contract with the National Aeronautics and Space Administration
This research has made use of NASA's Astrophysics Data System Bibliographic Services. Support for AMM is provided by NASA through Hubble Fellowship grant HST-HF2-51377 awarded by the Space Telescope Science Institute, which is operated by the Association of Universities for Research in Astronomy, Inc., for NASA, under contract NAS5-26555. Funding for the Sloan Digital Sky Survey IV has been provided by the Alfred P. Sloan Foundation, the U.S. Department of Energy Office of Science, and the Participating Institutions. SDSS acknowledges support and resources from the Center for High-Performance Computing at the University of Utah. The SDSS web site is www.sdss.org. 
SDSS is managed by the Astrophysical Research Consortium for the Participating Institutions of the SDSS Collaboration including the Brazilian Participation Group, the Carnegie Institution for Science, Carnegie Mellon University, the Chilean Participation Group, the French Participation Group, Harvard-Smithsonian Center for Astrophysics, Instituto de Astrofisica de Canarias, The Johns Hopkins University, Kavli Institute for the Physics and Mathematics of the Universe (IPMU) / University of Tokyo, Lawrence Berkeley National Laboratory, Leibniz Institut f\"{u}r Astrophysik Potsdam (AIP), Max-Planck-Institut f\"{u}r Astronomie (MPIA Heidelberg), Max-Planck-Institut f\"{u}r Astrophysik (MPA Garching), Max-Planck-Institut f\"{u}r Extraterrestrische Physik (MPE), National Astronomical Observatories of China, New Mexico State University, New York University, University of Notre Dame, Observatorio Nacional / MCTI, The Ohio State University, Pennsylvania State University, Shanghai Astronomical Observatory, United Kingdom Participation Group, Universidad Nacional Autonoma de Mexico, University of Arizona, University of Colorado Boulder, University of Oxford, University of Portsmouth, University of Utah, University of Virginia, University of Washington, University of Wisconsin, Vanderbilt University, and Yale University.

Facilities: \facility{SARA}, \facility{ATT}, \facility{Spitzer}, \facility{ESO}

\clearpage


\begin{thebibliography}{}
\bibitem[Auld et al.(2012)]{Auld} Auld, R., Smith, M.~W.~L., Bendo, G., et al.\ 2012, \mnras, 420, 1882 

\bibitem[Baldwin et al.(1981)]{BPT} Baldwin, J.~A., Phillips, M.~M., \& Terlevich, R.\ 1981, \pasp, 93, 5 


\bibitem[Blanco et al.(1975)]{Blanco75} Blanco, V.~M., Graham, J.~A., Lasker, B.~M., \& Osmer, P.~S.\ 1975, \apjl, 198, L63 

\bibitem[Brookes et al.(2006)]{Brookes} Brookes, M.~H., Lawrence, C.~R., Keene, J., et al.\ 2006, \apjl, 646, L41 

\bibitem[Charmandaris et al.(2000)]{Charmandaris} Charmandaris, V., Combes, F., \& van der Hulst, J.~M.\ 2000, \aap, 356, L1 

\bibitem[Crnojevi{\'c} et al.(2013)]{Crnojevic} Crnojevi{\'c}, D., Ferguson, A.~M.~N., Irwin, M.~J., et al.\ 2013, \mnras, 432, 832 

\bibitem[Crockett et al.(2012)]{Crockett} Crockett, R.~M., Shabala, S.~S., Kaviraj, S., et al.\ 2012, \mnras, 421, 1603 

\bibitem[Dopita et al.(2007)]{Dopita2007} Dopita, M., Hart, J., McGregor, P., et al.\ 2007, \apss, 310, 255 

\bibitem[Evans \& Koratkar(2004)]{EvansKoratkar} Evans, I.~N., \& Koratkar, A.~P.\ 2004, \apj, 617, 209 

\bibitem[Gil de Paz et al.(2005)]{GildePaz} Gil de Paz, A., Madore, B.~F., Boissier, S., et al.\ 2005, \apjl, 627, L29 

\bibitem[Graham \& Price(1981)]{GrahamPrice} Graham, J.~A., \& Price, R.~M.\ 1981, \apj, 247, 813 

\bibitem[Jacoby et al.(1984)]{JHC} Jacoby, G.~H., Hunter, D.~A., \& Christian, C.~A.\ 1984, \apjs, 56, 257 

\bibitem[Junkes et al.(1993)]{Junkes} Junkes, N., Haynes, R.~F., Harnett, J.~I., \& Jauncey, D.~L.\ 1993, \aap, 269, 29 

\bibitem[Keel(1989)]{Keel89} Keel, W.~C.\ 1989, European Southern Observatory Conference and Workshop Proceedings, 32, 427 

\bibitem[Keel et al.(2017)]{SARA} Keel, W.~C., Oswalt, T., Mack, P., et al.\ 2017, \pasp, 129, 015002 

\bibitem[Kennicutt et al.(2007)]{Kennicutt2007} Kennicutt, R.~C., Jr., Calzetti, D., Walter, F., et al.\ 2007, \apj, 671, 333 

\bibitem[Kewley et al.(2006)]{Kewley} Kewley, L.~J., Groves, B., Kauffmann, G., \& Heckman, T.\ 2006, \mnras, 372, 961 

\bibitem[Landolt(2009)]{Landolt} Landolt, A.~U.\ 2009, \aj, 137, 4186 

\bibitem[Lang et al.(2010)]{Lang} Lang, D., Hogg, D.~W., Mierle, K., Blanton, M., \& Roweis, S.\ 2010, \aj, 139, 1782 

\bibitem[McKinley et al.(2017)]{McKinley2017} McKinley, B., Tingay, S.~J., Carretti, E., et al.\ 2017, arXiv:1711.01751 

\bibitem[Morganti et al.(1991)]{Morganti91} Morganti, R., Robinson, A., Fosbury, R.~A.~E., et al.\ 1991, \mnras, 249, 91 

\bibitem[Neff et al.(2015)]{Neff2015} Neff, S.~G., Eilek, J.~A., \& Owen, F.~N.\ 2015, \apj, 802, 88 

\bibitem[Osmer(1978)]{Osmer78} Osmer, P.~S.\ 1978, \apjl, 226, L79 

\bibitem[Peterson et al.(1975)]{PDC75} Peterson, B.~A., Dickens, R.~J., \& Cannon, R.~D.\ 1975, Proceedings of the Astronomical Society of Australia, 2, 366 

\bibitem[Rejkuba et al.(2002)]{Rejkuba} Rejkuba, M., Minniti, D., Courbin, F., \& Silva, D.~R.\ 2002, \apj, 564, 688 

\bibitem[Santoro et al.(2015)]{Santoro2015} Santoro, F., Oonk, J.~B.~R., Morganti, R., \& Oosterloo, T.\ 2015, \aap, 574, A89 

\bibitem[Santoro et al.(2016)]{Santoro2016} Santoro, F., Oonk, J.~B.~R., Morganti, R., Oosterloo, T.~A., \& Tadhunter, C.\ 2016, \aap, 590, A37 

\bibitem[Schlafly \& Finkbeiner(2011)]{SchlaflyFinkbeiner} Schlafly, E.~F., \& Finkbeiner, D.~P.\ 2011, \apj, 737, 103 

\bibitem[Sternberg et al.(2003)]{Sternberg} Sternberg, A., Hoffmann, T.~L., \& Pauldrach, A.~W.~A.\ 2003, \apj, 599, 1333 

\bibitem[Sutherland et al.(1993)]{Sutherland} Sutherland, R.~S., Bicknell, G.~V., \& Dopita, M.~A.\ 1993, \apj, 414, 510 

\bibitem[Thomas et al.(2013)]{Thomas2013} Thomas, D., Steele, O., Maraston, C., et al.\ 2013, \mnras, 431, 1383

\bibitem[Tully et al.(2013)]{Tully2013} Tully, R.~B., Courtois, H.~M., Dolphin, A.~E., et al.\ 2013, \aj, 146, 86 

\bibitem[Tully et al.(2015)]{Tully2015} Tully, R.~B., Libeskind, N.~I., Karachentsev, I.~D., et al.\ 2015, \apjl, 802, L25 

\bibitem[Vacca et al.(1996)]{Vacca} Vacca, W.~D., Garmany, C.~D., \& Shull, J.~M.\ 1996, \apj, 460, 914 

\bibitem[Veilleux \& Osterbrock(1987)]{VO1987} Veilleux, S., \& Osterbrock, D.~E.\ 1987, \apjs, 63, 295 


\end{thebibliography}
\end{document}